# GaN Schottky diodes for proton beam monitoring

Jean-Yves Duboz*[1], Julie Zucchi[1], Eric Frayssinet[1], Patrick Chalbet[1], Sébastien Chenot[1], Maxime Hugues[1], Jean-Claude Grini[2], Richard Trimaud[2], Marie Vidal[2] and Joël Hérault[2]

*1 Université Côte d'Azur, CNRS, CRHEA, 06560, Valbonne, France*

*2-Université Côte d'Azur, Fédération Claude Lalanne;*

*Cyclotron Biomédical, Centre Antoine Lacassagne, Nice, France*

Email : jyd@crhea.cnrs.fr

Abstract

We have demonstrated that GaN Schottky diodes can be used for high energy (64.8 MeV) proton detection. Such proton beams are used for tumor treatment, for which accurate and radiation resistant detectors are needed. Schottky diodes have been measured to be highly sensitive to protons, to have a linear response with beam intensity and fast enough for the application. Some photoconductive gain was found in the diode leading to a good compromise between responsivity and response time. The imaging capability of GaN diodes in proton detection is also demonstrated.

AlGaInN alloys are nowadays widely used for LEDs for lighting and displays, for laser in the Blu-ray technology, and start to emerge for electronic applications in the radio frequency and power domains. GaN is a wide band gap material with a strong chemical and mechanical stability. The displacement energy is as large as 45 eV for the Ga atom and 109 eV for the N atom [1]. As a result, GaN is expected to be more robust than many other semiconductors against degradation under ionizing radiations. As an example, GaN based High Electron Mobility Transistors are expected to stand 10 times higher doses than their GaAs cousins, partly due to the difference in displacement energies and partly due to the piezoelectric field in nitrides [2]. Ion implantation in GaN shows degradation thresholds in the range from $10^{14}$ cm$^{-2}$ for heavy ions at an energy of 500 keV to $10^{16}$ cm$^{-2}$ for light ions at 10 keV [3]. The GaN device resistance against radiation has been studied experimentally. Apart a few studies on LEDs [4], most studies were devoted to Schottky diodes and transistors with the motivation of using GaN electronics for space applications. For neutron irradiation, a degradation threshold was found [5-7] for doses in the range of $10^{15}$ cm$^{-2}$. For proton irradiation, the threshold dose was found to be above $10^{13}$ cm$^{-2}$ for energies of 3 MeV [8] and degradation was clearly observed for a dose of $2\times10^{15}$ cm$^{-2}$ for energies of 5 MeV [9]. A more detailed analysis of structural defects created by an irradiation by 23 MeV protons at doses of $6\times10^{14}$ cm$^{-2}$ was reported [10,11]. As another benefice of its resistance to irradiation, GaN can be used for fabricating radiation hard particle detectors [12] in harsh environments such as synchrotrons [13] or for fabricating X-ray detectors for medical applications [14-16] . In the medical area, high energy particles are commonly used for cancer treatment. In particular, proton-therapy is used in cases where the tumor to be irradiated is close to vital and sensitive organs. A typical example is the eye tumor: The irradiation must be limited to the tumor in the eye while sparing whenever possible macula and optic nerve. For that, one takes advantage of a specific property of protons, which present very steep stopping profile in matter (Bragg peak) [17]. The proton dose has to be very well controlled, with an accuracy better than 3%, and detectors are thus needed both in vivo (to measure the dose during the patient irradiation) and on the beam (to monitor the beam prior to irradiation). Silicon detectors are used but present long term degradation. Wide band gap semiconductors have been proposed. Diamond is a good candidate due to its robustness and in vivo compatibility [18] but suffers from its industrial immaturity. On the contrary, GaN benefits from its large deployment for optics and electronics, and has been proposed for in vivo dose monitoring [19,20]. So far, only proton-luminescence has been tested in GaN. In this paper, we demonstrate that GaN can be used for a direct (electrical) reading detection of protons with an energy of 64.8 MeV, which are used for proton-therapy.

Schottky diodes have been fabricated on a GaN layer grown by metal organic vapour phase epitaxy. A 20 µm thick non-intentionally doped GaN layer was grown on a conductive GaN substrate (Lumilog). A TiAl ohmic contact was deposited on the back side first, and annealed at 750°C for 4 min. Then large Schottky contacts were deposited on the front side, with an area varying between 1 and 2 mm$^2$, and based on 10 nm of Pt followed by 100 nm of Au. A SiO$_2$ passivation layer was deposited next by Plasma Enhanced Chemical Vapor Deposition at 340°C. This deposition lasted 2 hours and also acted as an annealing step for the Schottky contact. A thick metal layer was finally deposited on the contact for facilitating the wire bonding. Samples with many diodes were mounted on ceramic chips with electrical connections. All measurements were made at room temperature.

Diodes were tested first outside the proton beam. We refer these measurements as "in the dark", although it was in the light of the room (we checked that the diode were insensitive to the light of the room). The diodes showed a rectifying behaviour, although not very strong, with some dispersion on the dark current ($I_d$) among diodes, between 0.2 and 10 nA at -2V. Then the diodes were measured in the proton beam of the MEDYCIC equipment, in the Lacassagne Proton-therapy Center [21]. The isochrone cyclotron delivers proton pulses with a duration of 7 ns and a frequency of 25 MHz. The beam is mono-energetic with a proton energy of 64.8 MeV. The fact that the beam is mono-energetic is crucial for proton-therapy. Indeed, protons lose energy when crossing matter, and the energy loss increases as their energy decreases. When protons are a few MeV energetic, their energy loss abruptly increases and they are stopped within less than 1 mm (Bragg peak), which makes protons so useful for tumour treatment. A multi-energetic beam would lose this property. The beam size on the detector position (upstream from the clinic treatment room) is about $1 \times 2$ cm$^2$. Although the proton beam is pulsed, all measurements are performed in CW as pulses are not resolved in our set up. The proton current can be varied from 10 pA to 100 nA. The total current in the beam ($I_t$) was measured for various proton currents and diode biases. It was found to be reproducible from diode to diode, within a factor of 2, and to be much larger than the dark current. The response to the proton (called protocurrent $I_p$, not to be confused with the proton current) is defined as the total current under the beam minus the dark current, $I_t$-$I_d$. Figure 1 shows, in log scale, the dark current and the protocurrent for one representative diode under a proton beam current of 20 nA. We first observe that the protocurrent follows the same bias dependence as the dark current. Second, at zero bias, the protocurrent is positive (0.47 µA). It changes sign between 0 and -0.1 V and then is negative for negative biases below -0.1V. Both observations indicate that the diode is not working in the normal photovoltaic mode. Indeed, the signal at zero bias should be negative, if it would originate from the Schottky depletion region only. This result can be explained as follows. The penetration depth of 64.8 MeV proton in GaN is a few mm, while the depletion region below the Schottky contact is few µm thick, the un-doped region is 20 µm tick and the substrate is 350 µm thick. Hence most of the absorption is in the doped substrate, where carriers rapidly recombine, as the field remains small even under an applied bias. We assume that the substrate contribution remains negligible except close to zero volt. A very small absorption occurs in the depletion region, leading to a small negative signal, and which should increase as the depletion region width, i.e. as $(\Phi-eV)^{1/2}$, where $\Phi$ is the Schottky barrier height. An intermediately large absorption occurs in the un-doped layer, where the field is close to zero at zero bias, but becomes non negligible under bias. This leads to a contribution which changes sign with bias. Electron external emission from the Schottky metal contact may also contribute and lead to a positive signal at zero bias. Finally, a photovoltaic effect on the back side of the sample could also be considered, leading to a positive signal at zero bias. The signal at zero bias results from all contributions, and turns out to be positive. Under an applied voltage, the contribution of the un-doped region dominates and leads to a photoconductive behaviour. In other words, the Schottky diode can be modelled as a rectifying contact in series with a resistance (un-doped region). Due to processing problems, the Schottky contact is not strongly rectifying and the current is mostly limited by the resistive layer, both in the dark and under the proton beam. This leads to a photoconductive behaviour. Please note that the same diode may have a photovoltaic behaviour under UV illumination as the absorption would be in the depletion layer only. Hence, the observed photoconductive behaviour in the proton beam is largely due to the large penetration depth of protons. When turning the beam on, we observed that the protocurrent was rapidly increasing to 90% of its final value, and then increasing with a few seconds to its final value. This slow transient is typical for a photoconductive behaviour. Similar behaviour was observed during turn-off. Note that the fast transient could not be observed with a resolution better than a fraction of second, so that it may also reveal a photoconductive behavior, although with a faster component than the observed slow transient. We will now discuss the absolute value of the protocurrent. Poly-methyl methacrylate (PMMA) is often used for proton dose calibration as its density is not too far the one of living tissues. The absorption depth in PMMA has been measured to be 29 mm. The density of GaN is 6.15 g/cm$^3$ while the PMMA one is 1.18g/cm$^3$, so that the absorption is stronger in GaN than in PMMA. The absorption depth calculated in GaN is supposed to be 8.6mm according to SRIM 2013 data tables [22]. For simplicity, we will take a uniform absorption in depth so that the absorption in a GaN layer of thickness W (in µm) is W/8600. The power deposited by the beam of section S in the diode of section s is W/8600×E×$I_p$×s/S, where E is the proton energy and $I_p$ the proton current. We then assume that protons obey the following rule of thumb: the energy needed to create an electron hole pair is three times the gap energy, i.e. about 10 eV. The charge created per second in the diode is then W/8600×E×Ip×s/S/(3×$E_g$). In a photovoltaic mode, without gain and with a unity collection efficiency, this charge created per second is equal to the protocurrent. Under a reverse bias of -2V, the depletion region W can be estimated to be about 4 µm. With s=1 mm$^2$ and s=2 cm$^2$, the calculation leads to a protocurrent of about 0.2µA. We have experimentally measured 100µA, which clearly shows that there is such photoconductive gain in the diode, confirming our assertion of a photoconductive behavior. The diode response has been measured as a function of the proton beam current, from 10 pA up to 100 nA. Figure 2 shows the result for various diode biases. We observe an excellent linearity over 5 decades.

One difficulty in the measurements was that the sample and sample holder became radioactive after exposure to the proton beam, and then impossible to handle. Hence, we tried to minimize the exposure time. In total, we can estimate that each sample remained at least 15 min in the beam, with an average current of 20 nA. This gives a cumulated dose larger than 0.5×10$^{14}$ cm$^{-2}$. We sometimes observed that the dark current in reverse bias increased after a beam exposure, in particular when it was originally very small, but returned to its initial value after some time (less than one hour) or after application of a positive bias. Hence, the dark current change was mostly due to the proto-generation of charges and their subsequent trapping. No real degradation could be observed from electrical characteristics, which is a first indication of the GaN resistance to ionizing radiations. Note that the radioactivity of GaN decreased rapidly. Various radioactive species are created, each of them with a different lifetime. Overall, the radioactivity returned to a low level (but not zero) after a couple of hours.

We have used the Schottky diodes to monitor the proton beam shape. The sample was mounted on a translation stage positioned in the center of the beam in the vertical direction (beam size about 2 cm in the vertical direction), and was moved in the horizontal direction over 5 cm (beam size on the order of 1 cm) across the beam. The signal was recorded as a function of position, thus giving the beam shape. Two diodes separated by about 3 mm on the same sample have been used. Figure 3 shows the result. The beam profile is slightly asymmetric, with a full width at half maximum of about 7.6 mm. The measured profile is the convolution of the actual beam profile and the detector width (1mm). The beam profile can be deduced from Fig.3 and is actually found to be close to 7.6mm. As an approximation, the convolution of two Gaussian profiles of width $W_1$ and $W_2$ gives a Gaussian profile with a width $W = \sqrt{W_1^2 + W_2^2}$. Hence, the beam profile is given by $\sqrt{7.6^2 - 1^2} = 7.53 mm$. A similar profile was obtained at 0V but with a slightly larger width of 8.2mm. This shows that the beam profile is convoluted with a larger detector width of approximately 3mm: This may indicate that at 0V, the signal is mainly arising from the back side of the sample, which is about 3 to 4 mm wide. Finally, the beam profile was measured with a Si diode (pin BP 104F) mounted so as to detect protons from the edge of the diode, corresponding to an effective width equal to the depletion zone, id est less than 10 µm. The full width at half maximum was found to be about 7.5mm, which confirms the measurements based on GaN Schottky diodes. The peak positions measured by the two diodes differ by about 3 mm, as expected from their separation on the sample. This shows that these Schottky diodes, if they are fabricated in an array, can be used for proton imaging. Typical sizes and periods for such an array of Schottky diodes would be about 100 µm for the proton-therapy application, and are easy to obtain from the technological point of view.

In conclusion, we have demonstrated that GaN Schottky diodes can be used to monitor proton beams at 64.8 MeV used for proton-therapy. Protons could be detected down to the smallest possible current, hence the sensitivity is high enough for the application. Diodes are linear in power, which is of prime importance for an accurate dosimetry. They are resistant to degradation up to a dose of at least $0.5 \times 10^{14}$ cm$^{-2}$. Time response is short enough (s) for the envisaged application. Some issues remain on elucidating the exact contribution of various parts of the device. Improvements in the processing and some changes in the epitaxial structure are likely to improve the device performance up to a commercial level. Such devices could be used for in vivo dosimetry as single detectors and also for external beam monitoring as arrays.

Acknowledgements
We acknowledge support from GANEX (ANR-11-LABX-0014). GANEX belongs to the public funded 'Investissements d'Avenir' program managed by the French ANR agency.

FIGURE CAPTIONS

FIGURE 1: Current versus voltage for a Schottky diode in the dark (black square) and under proton irradiation (proton beam current is 20 nA) (red triangle).

FIGURE 2: Current in the Schottky diode under proton irradiation versus proton beam intensity, for various detector biases

FIGURE 3: Normalized protocurrent in two Schottky diodes as a function of the translation stage position. The currents at a bias of -2V were -5.3 and -7.3 µA for diodes A and B respectively. Proton current is 10 nA.

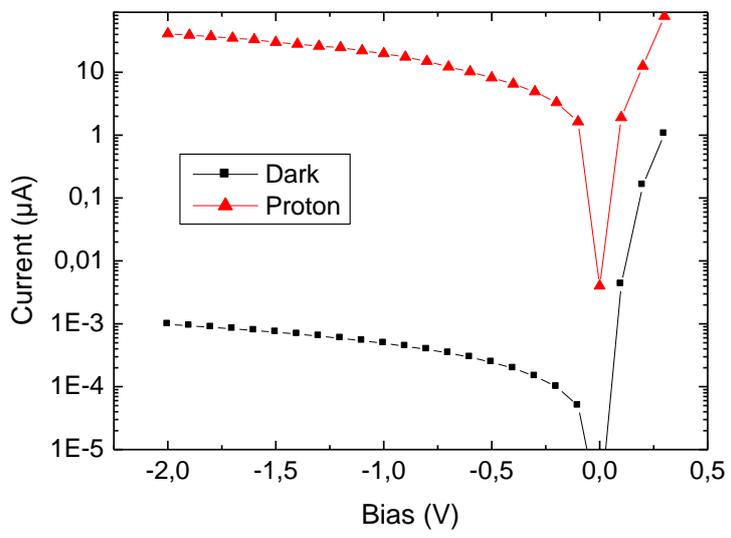

FIGURE 1

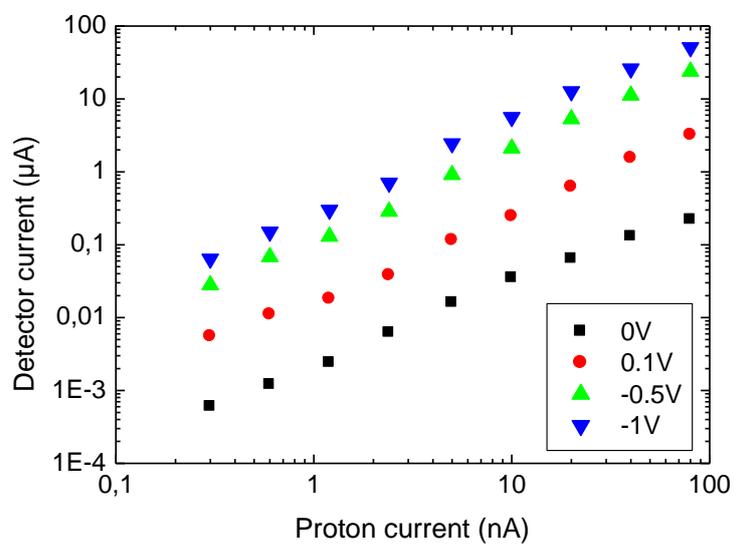

FIGURE 2

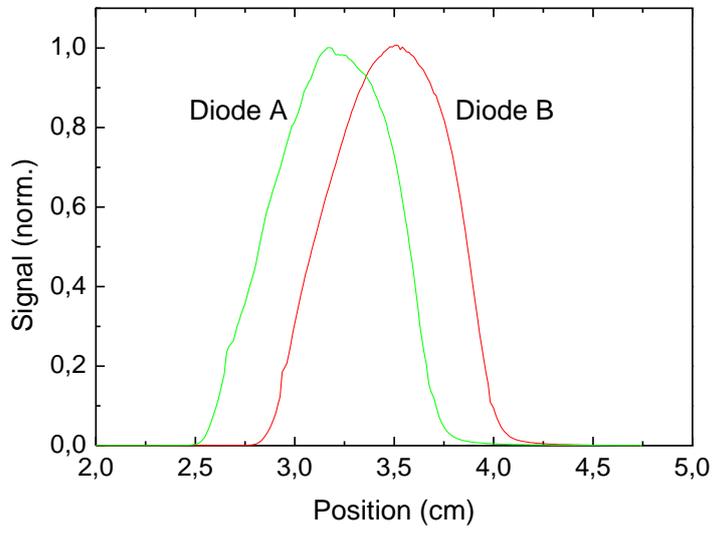

FIGURE 3